\magnification=1200\overfullrule=0pt\baselineskip=14.5pt
\vsize=23truecm \hsize=15.5truecm \overfullrule=0pt\pageno=0

\font\titlefont=cmbx10 scaled \magstep1
\font\sectnfont=cmbx8  scaled \magstep1
\def\mname{\ifcase\month\or January \or February \or March \or April
           \or May \or June \or July \or August \or September
           \or October \or November \or December \fi}
\def\date{\hbox{\strut\mname \number\year}}
\def\banner{\hfill\hbox{\vbox{\offinterlineskip\crnum}}\relax}
\def\manner{\hbox{\vbox{\offinterlineskip\crnum\date}}
               \hfill\relax}
\footline={\ifnum\pageno=0\manner\else\hfil\number\pageno\hfil\fi}
%
%
%
\newcount\FIGURENUMBER\FIGURENUMBER=0
\def\FIG#1{\expandafter\ifx\csname FG#1\endcsname\relax
               \global\advance\FIGURENUMBER by 1
               \expandafter\xdef\csname FG#1\endcsname
                              {\the\FIGURENUMBER}\fi}
\def\figtag#1{\expandafter\ifx\csname FG#1\endcsname\relax
               \global\advance\FIGURENUMBER by 1
               \expandafter\xdef\csname FG#1\endcsname
                              {\the\FIGURENUMBER}\fi
              \csname FG#1\endcsname\relax}
\def\fig#1{\expandafter\ifx\csname FG#1\endcsname\relax
               \global\advance\FIGURENUMBER by 1
               \expandafter\xdef\csname FG#1\endcsname
                      {\the\FIGURENUMBER}\fi
           Fig.~\csname FG#1\endcsname\relax}
\def\figand#1#2{\expandafter\ifx\csname FG#1\endcsname\relax
               \global\advance\FIGURENUMBER by 1
               \expandafter\xdef\csname FG#1\endcsname
                      {\the\FIGURENUMBER}\fi
           \expandafter\ifx\csname FG#2\endcsname\relax
               \global\advance\FIGURENUMBER by 1
               \expandafter\xdef\csname FG#2\endcsname
                      {\the\FIGURENUMBER}\fi
           figures \csname FG#1\endcsname\ and
                   \csname FG#2\endcsname\relax}
\def\figto#1#2{\expandafter\ifx\csname FG#1\endcsname\relax
               \global\advance\FIGURENUMBER by 1
               \expandafter\xdef\csname FG#1\endcsname
                      {\the\FIGURENUMBER}\fi
           \expandafter\ifx\csname FG#2\endcsname\relax
               \global\advance\FIGURENUMBER by 1
               \expandafter\xdef\csname FG#2\endcsname
                      {\the\FIGURENUMBER}\fi
           figures \csname FG#1\endcsname--\csname FG#2\endcsname\relax}
\newcount\TABLENUMBER\TABLENUMBER=0
\def\TABLE#1{\expandafter\ifx\csname TB#1\endcsname\relax
               \global\advance\TABLENUMBER by 1
               \expandafter\xdef\csname TB#1\endcsname
                          {\the\TABLENUMBER}\fi}
\def\tabletag#1{\expandafter\ifx\csname TB#1\endcsname\relax
               \global\advance\TABLENUMBER by 1
               \expandafter\xdef\csname TB#1\endcsname
                          {\the\TABLENUMBER}\fi
             \csname TB#1\endcsname\relax}
\def\table#1{\expandafter\ifx\csname TB#1\endcsname\relax
               \global\advance\TABLENUMBER by 1
               \expandafter\xdef\csname TB#1\endcsname{\the\TABLENUMBER}\fi
             Table \csname TB#1\endcsname\relax}
\def\tableand#1#2{\expandafter\ifx\csname TB#1\endcsname\relax
               \global\advance\TABLENUMBER by 1
               \expandafter\xdef\csname TB#1\endcsname{\the\TABLENUMBER}\fi
             \expandafter\ifx\csname TB#2\endcsname\relax
               \global\advance\TABLENUMBER by 1
               \expandafter\xdef\csname TB#2\endcsname{\the\TABLENUMBER}\fi
             Tables \csname TB#1\endcsname{} and
                    \csname TB#2\endcsname\relax}
\def\tableto#1#2{\expandafter\ifx\csname TB#1\endcsname\relax
               \global\advance\TABLENUMBER by 1
               \expandafter\xdef\csname TB#1\endcsname{\the\TABLENUMBER}\fi
             \expandafter\ifx\csname TB#2\endcsname\relax
               \global\advance\TABLENUMBER by 1
               \expandafter\xdef\csname TB#2\endcsname{\the\TABLENUMBER}\fi
            Tables \csname TB#1\endcsname--\csname TB#2\endcsname\relax}
\newcount\REFERENCENUMBER\REFERENCENUMBER=0
\def\REF#1{\expandafter\ifx\csname RF#1\endcsname\relax
               \global\advance\REFERENCENUMBER by 1
               \expandafter\xdef\csname RF#1\endcsname
                         {\the\REFERENCENUMBER}\fi}
\def\reftag#1{\expandafter\ifx\csname RF#1\endcsname\relax
               \global\advance\REFERENCENUMBER by 1
               \expandafter\xdef\csname RF#1\endcsname
                      {\the\REFERENCENUMBER}\fi
             \csname RF#1\endcsname\relax}
\def\ref#1{\expandafter\ifx\csname RF#1\endcsname\relax
               \global\advance\REFERENCENUMBER by 1
               \expandafter\xdef\csname RF#1\endcsname
                      {\the\REFERENCENUMBER}\fi
             [\csname RF#1\endcsname]\relax}
\def\refto#1#2{\expandafter\ifx\csname RF#1\endcsname\relax
               \global\advance\REFERENCENUMBER by 1
               \expandafter\xdef\csname RF#1\endcsname
                      {\the\REFERENCENUMBER}\fi
           \expandafter\ifx\csname RF#2\endcsname\relax
               \global\advance\REFERENCENUMBER by 1
               \expandafter\xdef\csname RF#2\endcsname
                      {\the\REFERENCENUMBER}\fi
             [\csname RF#1\endcsname--\csname RF#2\endcsname]\relax}
\def\refand#1#2{\expandafter\ifx\csname RF#1\endcsname\relax
               \global\advance\REFERENCENUMBER by 1
               \expandafter\xdef\csname RF#1\endcsname
                      {\the\REFERENCENUMBER}\fi
           \expandafter\ifx\csname RF#2\endcsname\relax
               \global\advance\REFERENCENUMBER by 1
               \expandafter\xdef\csname RF#2\endcsname
                      {\the\REFERENCENUMBER}\fi
            [\csname RF#1\endcsname,\csname RF#2\endcsname]\relax}
\def\refs#1#2{\expandafter\ifx\csname RF#1\endcsname\relax
               \global\advance\REFERENCENUMBER by 1
               \expandafter\xdef\csname RF#1\endcsname
                      {\the\REFERENCENUMBER}\fi
           \expandafter\ifx\csname RF#2\endcsname\relax
               \global\advance\REFERENCENUMBER by 1
               \expandafter\xdef\csname RF#2\endcsname
                      {\the\REFERENCENUMBER}\fi
            [\csname RF#1\endcsname,\csname RF#2\endcsname]\relax}
\def\refss#1#2#3{\expandafter\ifx\csname RF#1\endcsname\relax
               \global\advance\REFERENCENUMBER by 1
               \expandafter\xdef\csname RF#1\endcsname
                      {\the\REFERENCENUMBER}\fi
           \expandafter\ifx\csname RF#2\endcsname\relax
               \global\advance\REFERENCENUMBER by 1
               \expandafter\xdef\csname RF#2\endcsname
                      {\the\REFERENCENUMBER}\fi
           \expandafter\ifx\csname RF#3\endcsname\relax
               \global\advance\REFERENCENUMBER by 1
               \expandafter\xdef\csname RF#3\endcsname
                      {\the\REFERENCENUMBER}\fi
            [\csname RF#1\endcsname,\csname
                RF#2\endcsname,\csname RF#3\endcsname]\relax}
\def\Ref#1{\expandafter\ifx\csname RF#1\endcsname\relax
               \global\advance\REFERENCENUMBER by 1
               \expandafter\xdef\csname RF#1\endcsname
                      {\the\REFERENCENUMBER}\fi
             Ref.~\csname RF#1\endcsname\relax}
\def\Refs#1#2{\expandafter\ifx\csname RF#1\endcsname\relax
               \global\advance\REFERENCENUMBER by 1
               \expandafter\xdef\csname RF#1\endcsname
                      {\the\REFERENCENUMBER}\fi
           \expandafter\ifx\csname RF#2\endcsname\relax
               \global\advance\REFERENCENUMBER by 1
               \expandafter\xdef\csname RF#2\endcsname
                      {\the\REFERENCENUMBER}\fi
        Refs.~\csname RF#1\endcsname{},\csname RF#2\endcsname\relax}
\def\Refto#1#2{\expandafter\ifx\csname RF#1\endcsname\relax
               \global\advance\REFERENCENUMBER by 1
               \expandafter\xdef\csname RF#1\endcsname
                      {\the\REFERENCENUMBER}\fi
           \expandafter\ifx\csname RF#2\endcsname\relax
               \global\advance\REFERENCENUMBER by 1
               \expandafter\xdef\csname RF#2\endcsname
                      {\the\REFERENCENUMBER}\fi
            Refs.~\csname RF#1\endcsname--\csname RF#2\endcsname]\relax}
\def\Refand#1#2{\expandafter\ifx\csname RF#1\endcsname\relax
               \global\advance\REFERENCENUMBER by 1
               \expandafter\xdef\csname RF#1\endcsname
                      {\the\REFERENCENUMBER}\fi
           \expandafter\ifx\csname RF#2\endcsname\relax
               \global\advance\REFERENCENUMBER by 1
               \expandafter\xdef\csname RF#2\endcsname
                      {\the\REFERENCENUMBER}\fi
        Refs.~\csname RF#1\endcsname{} and \csname RF#2\endcsname\relax}
\newcount\EQUATIONNUMBER\EQUATIONNUMBER=0
\def\EQ#1{\expandafter\ifx\csname EQ#1\endcsname\relax
               \global\advance\EQUATIONNUMBER by 1
               \expandafter\xdef\csname EQ#1\endcsname
                          {\the\EQUATIONNUMBER}\fi}
\def\eqtag#1{\expandafter\ifx\csname EQ#1\endcsname\relax
               \global\advance\EQUATIONNUMBER by 1
               \expandafter\xdef\csname EQ#1\endcsname
                      {\the\EQUATIONNUMBER}\fi
            \csname EQ#1\endcsname\relax}
\def\EQNO#1{\expandafter\ifx\csname EQ#1\endcsname\relax
               \global\advance\EQUATIONNUMBER by 1
               \expandafter\xdef\csname EQ#1\endcsname
                      {\the\EQUATIONNUMBER}\fi
            \eqno(\csname EQ#1\endcsname)\relax}
\def\EQNM#1{\expandafter\ifx\csname EQ#1\endcsname\relax
               \global\advance\EQUATIONNUMBER by 1
               \expandafter\xdef\csname EQ#1\endcsname
                      {\the\EQUATIONNUMBER}\fi
            (\csname EQ#1\endcsname)\relax}
\def\eq#1{\expandafter\ifx\csname EQ#1\endcsname\relax
               \global\advance\EQUATIONNUMBER by 1
               \expandafter\xdef\csname EQ#1\endcsname
                      {\the\EQUATIONNUMBER}\fi
          Eq.~(\csname EQ#1\endcsname)\relax}
\def\eqand#1#2{\expandafter\ifx\csname EQ#1\endcsname\relax
               \global\advance\EQUATIONNUMBER by 1
               \expandafter\xdef\csname EQ#1\endcsname
                        {\the\EQUATIONNUMBER}\fi
          \expandafter\ifx\csname EQ#2\endcsname\relax
               \global\advance\EQUATIONNUMBER by 1
               \expandafter\xdef\csname EQ#2\endcsname
                      {\the\EQUATIONNUMBER}\fi
         Eqs.~(\csname EQ#1\endcsname) and (\csname EQ#2\endcsname)\relax}
\def\eqto#1#2{\expandafter\ifx\csname EQ#1\endcsname\relax
               \global\advance\EQUATIONNUMBER by 1
               \expandafter\xdef\csname EQ#1\endcsname
                      {\the\EQUATIONNUMBER}\fi
          \expandafter\ifx\csname EQ#2\endcsname\relax
               \global\advance\EQUATIONNUMBER by 1
               \expandafter\xdef\csname EQ#2\endcsname
                      {\the\EQUATIONNUMBER}\fi
          Eqs.~\csname EQ#1\endcsname--\csname EQ#2\endcsname\relax}
\def\APEQNO#1{\expandafter\ifx\csname EQ#1\endcsname\relax
               \global\advance\EQUATIONNUMBER by 1
               \expandafter\xdef\csname EQ#1\endcsname
                      {\the\EQUATIONNUMBER}\fi
            \eqno(\APPENDIXNUMBER.\csname EQ#1\endcsname)\relax}
\def\APEQNM#1{\expandafter\ifx\csname EQ#1\endcsname\relax
               \global\advance\EQUATIONNUMBER by 1
               \expandafter\xdef\csname EQ#1\endcsname
                      {\the\EQUATIONNUMBER}\fi
            (\APPENDIXNUMBER.\csname EQ#1\endcsname)\relax}
\def\apeq#1{\expandafter\ifx\csname EQ#1\endcsname\relax
               \global\advance\EQUATIONNUMBER by 1
               \expandafter\xdef\csname EQ#1\endcsname
                      {\the\EQUATIONNUMBER}\fi
          Eq.~(\APPENDIXNUMBER.\csname EQ#1\endcsname)\relax}
%
\newcount\SECTIONNUMBER\SECTIONNUMBER=0
\newcount\SUBSECTIONNUMBER\SUBSECTIONNUMBER=0
\def\appendix#1#2{\global\let\APPENDIXNUMBER=#1\bigskip\goodbreak
     \line{{\sectnfont Appendix \APPENDIXNUMBER.\ #2}\hfil}\smallskip}
\def\section#1{\global\advance\SECTIONNUMBER by 1\SUBSECTIONNUMBER=0
      \bigskip\goodbreak\line{{\sectnfont \the\SECTIONNUMBER.\ #1}\hfil}
      \smallskip}
\def\subsection#1{\global\advance\SUBSECTIONNUMBER by 1
      \bigskip\goodbreak\line{{\sectnfont
         \the\SECTIONNUMBER.\the\SUBSECTIONNUMBER.\ #1}\hfil}
      \smallskip}
%
\def\AnnP{{\sl Ann.\ Phys.\ }}

\def\NP{{\sl Nucl.\ Phys.\ }}

\def\PR{{\sl Phys.\ Rev.\ }}

\def\crnum{\hbox{HLRZ 20/93 \strut}}

\def\banner{\hfill\hbox{\vbox{\crnum}}\relax}
\def\manner{\hbox{\vbox{\offinterlineskip\bigskip\bigskip
                        \crnum\date}}\hfill\relax}
\newdimen\digitwidth\setbox0=\hbox{\rm0}\digitwidth=\wd0

  \def\L{{\scriptscriptstyle L}}

\def\lb{\hfil\penalty-10000}

\def\ie{{\sl i.e.\/}}

{\vsize=20truecm\banner\bigskip\bigskip\bigskip\baselineskip=15pt
\begingroup\titlefont\obeylines
\hfil CRITICAL HYSTERESIS\hfil
\endgroup\bigskip
\bigskip\centerline{Sourendu Gupta}
\centerline{HLRZ, c/o KFA J\"ulich, D-5170 J\"ulich, Germany.}
\bigskip\bigskip\bigskip\bigskip
\centerline{\bf ABSTRACT}\medskip
Hysteresis is observed at second order phase transitions. Universal scaling
formul\ae{} for the areas of hysteresis loops are written down. Critical
exponents are defined, and related to other exponents for static and dynamic
critical phenomena. These relations are verified with Langevin dynamics in
both the critical and tricritical mean-field models. A finite-size scaling
relation is tested in the two-dimensional Ising model with heat-bath dynamics.
\vfil\eject}

Hysteresis has been studied quantitatively for one hundred years now
\ref{ewing}. This phenomenon is well-known at first-order phase transitions.
We find that it can also be observed at a second-order transition. Such
critical hysteresis is characterised by certain scaling laws, \ie, universal
exponents and scaling functions. The exponents are related to well-known
static and dynamical critical exponents. Critical hysteresis is thus seen
to be entirely consistent with the phenomenon of critical slowing down.
Furthermore, when the critical point is a limit of a line of first order
phase transitions, the scaling function for the hysteresis loop area is a
smooth limit of those obtained along the line of transitions.

The characteristic observation in hysteresis is that when a system is
subjected to a cyclic external field, $h$, its response, $m$, lags behind
the field. Thus the curve $m(h)$ forms a loop.
Although many different kinds of systems show hysteretic behaviour,
in this letter we shall use a language appropriate to magnetic systems.
Thus, the external field will be a periodic magnetic
field, with period $T=2\pi/\omega$, and an amplitude $h_0$. The
response of the system will be characterised by an induced time-dependent
magnetisation. All the statements we shall make have their analogues
for other kinds of phase transitions, provided the external field is
linearly coupled to the order parameter, which forms the response.

Of interest in most experiments is the area of the hysteresis loop, defined as
$$A(\omega,h_0)\;=\;{1\over4\pi}\oint m(t) dh,      \EQNO{area}$$
where the integral is over one period of the applied field. This is the
energy dissipated by the system. Also of interest are the shapes of the
hysteresis loops. There have been many early studies of the dependence
of $A$ on $\omega$ and $h_0$ \ref{mad}. Recent theoretical work has focussed
on the low-temperature phase of $O(N)$ models\refs{mad}{deepak}. In these
models power laws have been derived for the scaling of hysteresis loop areas
with $h_0$ and $\omega$.

The essential physics of hysteresis is the lag of the response behind the
force. This is due to finite autocorrelation times.
Whenever the inverse frequency of
the external force is comparable to an appropriate
autocorrelation time of the system, one expects to see hysteresis. Due to
the phenomenon of critical slowing down, one must therefore see hysteresis
also at a second-order phase transition. We shall first outline the dynamics
that we use. Next, assuming the existence of this phenomenon, we shall
present scaling laws for the loop area and define critical exponents. Finally
we shall present some examples of critical hysteresis.
The scaling laws are seen to be satisfied with appropriate exponents.
Further details will be presented elsewhere \ref{future}.

In this letter we shall take the dynamics to be
given by a Langevin equation without conservation laws. We shall consider
real one-component fields $\phi_i$ on the sites, $i$, of
a $d$-dimensional lattice, whose interactions
are specified by a Hamiltonian $H$. In terms of a Gaussian white noise
$\nu_i(t)$, with width $\Gamma$, the Langevin equation can be written as
$${d\phi_i\over dt}\;=\; \Gamma {\delta H\over\delta\phi_i} +\nu_i(t).
       \EQNO{lange}$$
Note that this dynamics has no conserved quantities.
We shall restrict our attention to the zero mode or magnetisation,
$\phi=\sum_i\phi_i$. The argument is that this Fourier mode has the
largest autocorrelation time, and therefore the long-time behaviour
of the system is due to this.
It is known \refs{msr}{zinn} that all correlations functions arising
from this one-mode dynamics are exactly equal to those obtained from
the Euclidean quantum mechanics specified by the Euclidean action
$$S\;=\;{1\over\hbar} \int dt\int \left\{
       {1\over2}\partial_t\phi^2 + {1\over2} W^2 -
          {\hbar\over2} W'\right\},         \EQNO{susy}$$
where $W$ stands for $\Gamma\delta H/\delta\phi$, and the prime
denotes a derivative with respect to $\phi$. The `Planck's constant',
$\hbar$, is equal to $\Gamma/L^d$.
The infinite volume limit of the theory is therefore obtained as the
`classical limit'. This is investigated by the Euler-Lagrange equations
arising from the Euclidean action above, and reduces to the Langevin
equation with the noise term dropped. This method has previously been
used \ref{zinn} to obtain the instanton solution in the dynamics.

For technical reasons, we shall assume that the amplitude of the applied
magnetic field is small and linear response theory is applicable.
A flucutuation-dissipation theorem \ref{sushanta} relates the energy
dissipation to the the absorptive part of the complex susceptibility,
$\chi''(\omega)$, through the relation
$$A(\omega,h_0)\;=\;{1\over2}h_0^2\omega\chi''(\omega).  \EQNO{fd}$$
Along with the definition
$$\chi''(\omega)\;=\;{\cal I}{\rm m} \int_0^\infty
       dt {\rm e}^{-i\omega t} \langle\phi(t)\phi(0)\rangle,
           \EQNO{defchi}$$
where the angular brackets denote averaging over the equilibrium ensemble,
we now have the technical machinery to compute quantities of interest.
Note that the average over the equilibrium ensemble in the formula
above is a reflection of the fact that we are working in linear response
theory.

As an example, consider the most trivial of models--- a Gaussian integral.
The Hamiltonian is
$$H\;=\; {1\over2} \omega_0^2 \phi^2.      \EQNO{gauss}$$
By a variety of methods one can construct the autocorrelation function in
Langevin dynamics
$$C(t)\;=\; {\rm e}^{-t/\tau}, \qquad\qquad(\tau=1/\omega_0^2). \EQNO{acor}$$
This yields hysteresis loop areas
$$A(\omega,h_0)\;=\; {1\over2} {h_0^2\omega\tau^2\over1+\omega^2\tau^2}.
                      \EQNO{gaussr}$$
The claim made earlier, that hysteresis vanishes only when the autocorrelation
time does, is explicitly demonstrated in this model. Note also the $1/\omega$
fall at large $\omega$. This behaviour has been observed, for example,
in $O(N)$ models in dimensions $d>1$ \ref{deepak}. It will be seen again in
critical hysteresis. Such behaviour is obtained whenever a pole in the
spectral density determines the behaviour of $A$ at large frequencies.

Now we write a scaling relation at second-order phase transitions for
the hysteresis loop area as a function of $\omega$ and $h_0$. The necessary
presence of an external magnetic field implies that we can tune
$\theta=(T-T_c)/T_c$ to zero, and let $h_0$ control the approach to
criticality. As a result, exponents corresponding to $\theta$ do not
appear in the scaling relations. We write
$$A(\omega,h_0,\theta)\;=\;h_0^\lambda F(\omega h_0^\mu).   \EQNO{scaling}$$
In the limit $\omega\to0$, this must yield the usual relation
between the magnetic field and the magnetisation. This implies
$$\lambda\;=\; 1+{1\over\delta}.      \EQNO{lambda}$$
Time-dependent phenomena are characterised by the scaling of the field
autocorrelation time $\tau$ with increasing correlation length $\xi$. The
relation between these two quantities, $\tau\sim\xi^z$, defines a dynamical
critical exponent $z$. As one approaches the critical point by tuning
$h_0$, the scaling of $\omega$
required to keep the hysteresis loop area fixed must be given by the
relation between $\tau$ and $\xi$ above. This gives the remaining
critical exponent
$$\mu\;=\;-{z\nu\over\beta\delta}  \EQNO{mu}$$
The scaled variable $\omega h_0^\mu$ shall be called $\Omega$.
For non-zero $\theta$, there will be, of course, corrections to the
scaling form given above which become apparent as $h_0\to0$.

A trivial illustration of the scaling formula is provided by the
Gaussian integral analysed before. Coupling a magnetic field linearly to
$\phi$, it is easy to see that $\delta=1$ and, since $\omega_0$ is
independent of $h$, $\nu/\delta\beta=0$. As a result, $\lambda=2$,
consistent with the overall factor of $h_0^2$, and $\mu=0$, consistent
with the fact that the only $h_0$ dependence is in this factor.

The scaling laws illuminate the origin of critical hysteresis.
Provided that the scaling function $F(\Omega)$ is non-zero,
one must have critical hysteresis.
This is a consequence of critical slowing down.
We anticipate the subsequent discussion leading up to the fact that
$F(\Omega)$ has a maximum at (say) $\Omega_0$,
decays for $\Omega>\Omega_0$ as $1/\Omega$,
and for small $\Omega$ increases as a power of $\Omega$ (see \fig{tdlg}).
As a consequence, the scaling form above predicts that as $h_0$ decreases,
the hysteresis loop area increases at any fixed arbitrarily small frequency.
At large frequencies, on the other hand, the system cannot follow changes
in the driving field, and the magnetisation remains zero.
In this qualitative sense, hysteretic behaviour at a critical point
closely resembles that at a first order transition.

The scaling relations for a thermodynamically large system imply, in the
usual way, finite-size scaling (FSS) relations. Well-known arguments
\ref{barber} can be adapted to obtain various scaling formul\ae. Of
particular interest is the finite-size scaling of the frequency, $\omega_m$,
at which the hysteresis loop area reaches its maximum. When systems of
different sizes, $L$, are studied, each at some appropriately defined
pseudo-critical coupling, and $h_0$ is independent of $L$ but smaller than
the value at which finite-size rounding sets in for the largest system,
then
$$\omega_m(L)\;\sim\; L^z.  \EQNO{fss}$$
Note that this relation gives a method for the measurement of $z$.
The usual FSS methods for dynamics only measure the combination
of exponents $z/\nu$.

Our first illustrative example is of the time-dependent Landau-Ginzburg model
at its critical point. The model is defined by the Langevin
equation in \eq{lange} along with the zero-mode Hamiltonian
$$H\;=\;{r\over2}\phi^2 +{g\over4!}\phi^4 - h\phi.
         \EQNO{gl}$$
A (possibly time-dependent) magnetic field $h$ has been introduced. The
critical theory corresponds to $r=0$. We investigate this theory for large
volume systems. As explained earlier, this requires the solution of the
ordinary differential equation obtained by dropping the noise term in the
Langevin equation. After the scalings
$$t\to a^2t,\quad\phi\to a\phi,\quad h\to h/a,\qquad\qquad
a=\left({6\over\Gamma g}\right)^{1/4}, \EQNO{redef}$$
we obtain the equation
$${d\phi\over dt}\;=\; -\phi^3 - c\theta\phi + h(t).  \EQNO{tdgl}$$
We used a harmonic form, $h_0\cos\omega t$, for the driving field $h(t)$.
Note that the scaling of $t$ implies one for the frequency $\omega$.
We have assumed that $r\sim\theta$, and that $c$ includes a factor
$\sqrt{6\Gamma/g}$, whose temperature-dependence can be neglected for
sufficiently small $\theta$. The presence of hysteresis makes this a stiff
equation. It was solved using Gear's method, starting from an arbitrary
initial value and integrating until transients died out to the precision
of the arithmetic. The subsequent time history of the system gave the stable
limit cycles corresponding to hysteresis. Scaling was checked by varying
$h_0$ over three decades and $\omega$ over four.

The solutions showed hysteretic behaviour quite clearly. At very high
frequencies, the system could not follow the applied magnetic field; and
the magnetisation remained zero and time independent. As $\omega$ decreased,
the hysteresis loop $m(h)$ opened into an ellipse with its major axis along
$h$. With further decrease in $\omega$, the major axis turned, and the
ellipse began to get dented, until it resembled familiar hysteresis loops.
Eventually, with decreasing $\omega$ these loops became thinner.
We calculated the areas of the hysteresis loops as a function of $\omega$
and $h_0$ and found that the scaling law given in \eq{scaling} is obtained
with high precision. We found the exponents
$$\lambda={4\over3}, \qquad{\rm and}\qquad \mu=-{2\over3}.  \EQNO{tdlgexp}$$
These are completely consistent with the values $\delta=3$, $\nu=\beta=1/2$
and $z=2$. The scaling function can be identified by our
computation. It is shown in \fig{tdlg}. It has a single maximum at
$\Omega\approx0.93$.

These results were obtained for $c\theta=0$.
Positive values for this parameter correspond to approaching the critical
point from the disordered side. As the value of this parameter increased,
distortions in $F(\Omega)$ could be seen at higher and higher $\Omega$.
For small $\Omega$ these tended to decrease $F$, and push $\Omega_0$
towards larger values. For $c\theta$ less than zero, we approach criticality
from the ordered side, where there are coexisting phases. Varying $c\theta$
in this region, for any fixed $h_0$, we found that $A(\omega,h_0)$ approached
its value at criticality smoothly.

It is interesting that the hysteresis loops themselves are scaling functions.
At $\theta=0$ it was found that
$$m(h;\omega,h_0)\;=\;
     h_0^{1/\delta} L(h/h_0;\Omega),   \EQNO{loopshape}$$
where the loop shape $L$ depended on $\Omega$.
This is a form of a dynamical principle of corresponding states.
The shapes of the loops are shown in \fig{tdlg} at the
appropriate points below the scaled frequency axis.
For $\Omega\to\infty$, $L=0$ identically,
and for $\Omega\to0$, $L(x)=x^{1/\delta}$.

The exponents and scaling functions enjoy a high degree of universality.
We checked this by changing the form of the cyclic driving field $h(t)$.
For a variety of forms, all with time-reversal invariance, we found the
same values for the exponents. Whether we used triangular waves, or
piecewise constant functions did not matter. Furthermore, the scaling
function obtained for each shape of the driving field was simply related
to that obtained for the harmonic form. We found that all the scaling
functions mapped on to each other with multiplicative redefinitions
of $h_0$ and $\omega$. This leads to the form $c_1F(c_2\Omega)$,
with universal $F(\Omega)$,
where $c_1$ and $c_2$ are constants independent of $h_0$, $\omega$ and
$\theta$, dependent only on the shape of the driving force.

A second example is provided by the Langevin dynamics of the $\phi^6$
model at its tricritical point. After appropriate scalings, we have, for
the large-volume limit of the Langevin equation
$${d\phi\over dt}\;=\; -\phi^5 + h(t).  \EQNO{tric}$$
Numerical solutions were obtained for this equation, varying $h_0$ and
$\omega$ as before. The critical exponents are well-known \ref{tric}---
$\delta=5$, $\beta=1/4$, $\nu=1/2$ and $z=2$. These yield $\lambda=6/5$
and $\mu=-4/5$. These described the numerical results extremely well.
The universal scaling function thus obtained is shown in \fig{tdlg}.
As for the $\phi^4$ model, this function is universal over cyclic
driving fields. It is also interesting that the scaling functions for
critical and tricritical dynamics are identical for $\Omega\gg\Omega_m$.
Universal hysteresis loop shapes were also observed, and found to be
qualitatively similiar to those for the $\phi^4$ theory.

We checked the FSS relation in \eq{fss} through heat-bath simulations
of the two-dimensional Ising model. We simulated $32^2$, $40^2$ and
$64^2$ lattices at the pseudo-critical coupling, $\beta_c^\L$, defined
by the maximum of the specific heat.
Equlibration at zero field was achieved by $10^6$ sweeps with the
Swendson-Wang cluster algorithm. Hysteresis runs were performed changing
the external field cyclically between $\pm h_0$ in steps of $\Delta h$.
On all the lattices we chose $h_0=8\times10^{-4}$ and $\Delta h=10^{-4}$.
At each value of the field we ran $N$ steps of the heat-bath algorithm.
The frequency is given by $\omega=2\pi\Delta h/Nh_0$. Expectation values
and errors for hysteresis loop areas were obtained by jack-knifing
measurements from 500 loop traversals at each $\omega$ and $L$ into
10 blocks. A series of runs on each size of lattice located $\omega_m$
with fairly high precision. The data are shown in in \fig{ising} along
with the scaling law obtained using the dynamical exponent $z=2$ for
heat-bath dynamics. The hysteresis loop shapes were seen to be universal,
and qualitatively similiar to the time-dependent Landau-Ginzburg model.

In conclusion, we have observed hysteresis at a critical point. Scaling
relations and critical exponents for hysteresis loop areas have been
identified and related to other static and dynamic critical exponents.
Finite-size scaling relations following from these have been written down.
Since the observation of hysteresis is no guarantee that a first-order
phase transition is seen, these relations are important in deciding the
nature of the transition. Scaled hysteresis loop shapes and areas are also
universal, characterised by functions which do not depend on the precise
form of the driving force. The scaling functions identified here should
be interesting objects of study, since they contain information on the
complex structure of the underlying spectral density in an easily accessible
fashion.
\vfil\eject
\centerline{\sectnfont REFERENCES}\bigskip
\item{\reftag{ewing})}
   J.~A.~Ewing and H.~G.~Klaassen, {\sl Phil.\ Trans.\ R.\ Soc.\ London\/},
   Series A, 184 A (1893) 985.
\item{\reftag{mad})}
   See M.~Rao, H.~R.~Krishnamurthy and R.~Pandit, \PR B 42 (1990) 856,
   for a complete list of references.
\item{\reftag{deepak})}
   D.~Dhar and P.~B.~Thomas, {\sl J.\ Phys.\/}, A 25 (1992) 4967.
\item{\reftag{future})}
   S.~Gupta, in preparation.
\item{\reftag{msr})}
   P.~C.~Martin, E.~D.~Siggia and H.~A.~Rose, \PR A 8 (1973) 423.
\item{\reftag{zinn})}
   F.~Cooper and B.~Freedman, \AnnP 146 (1983) 262;\lb
   J.~C.~Niel and J.~Zinn-Justin, \NP B280 [FS18] (1987) 355.
\item{\reftag{sushanta})}
   See, for example,
   P.~C.~Martin, {\sl `Measurements and Correlation Functions'\/},
   1968, Gordon and Breach, New York.
\item{\reftag{barber})}
   M.~N.~Barber, {\sl `Phase Transitions and Critical Phenomena'\/}, Vol 8,
   p.\ 145, Eds.\ C.~Domb and J.~L.~Lebowitz, 1983, Academic Press, London.
\item{\reftag{tric})}
   I.~D.~Lawrie and S.~Sarbach, {\sl `Phase Transitions and Critical
   Phenomena'\/}, Vol 9,  p.\ 1, Eds.\ C.~Domb and J.~L.~Lebowitz, 1984,
   Academic Press, London.
\vfil\eject

\midinsert\vskip13truecm
\centerline{FIGURE \figtag{tdlg}}
\includegraphics{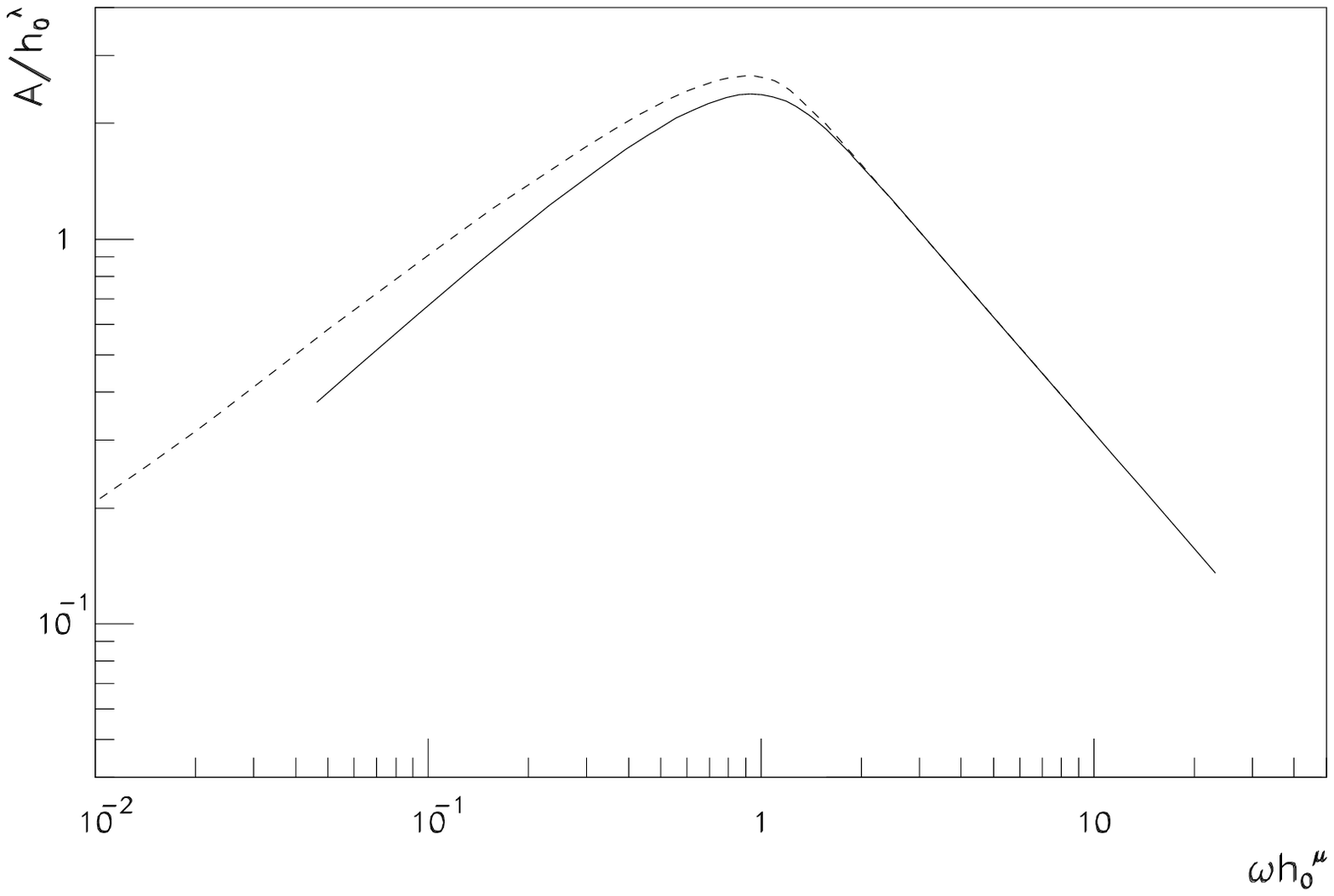}
\includegraphics{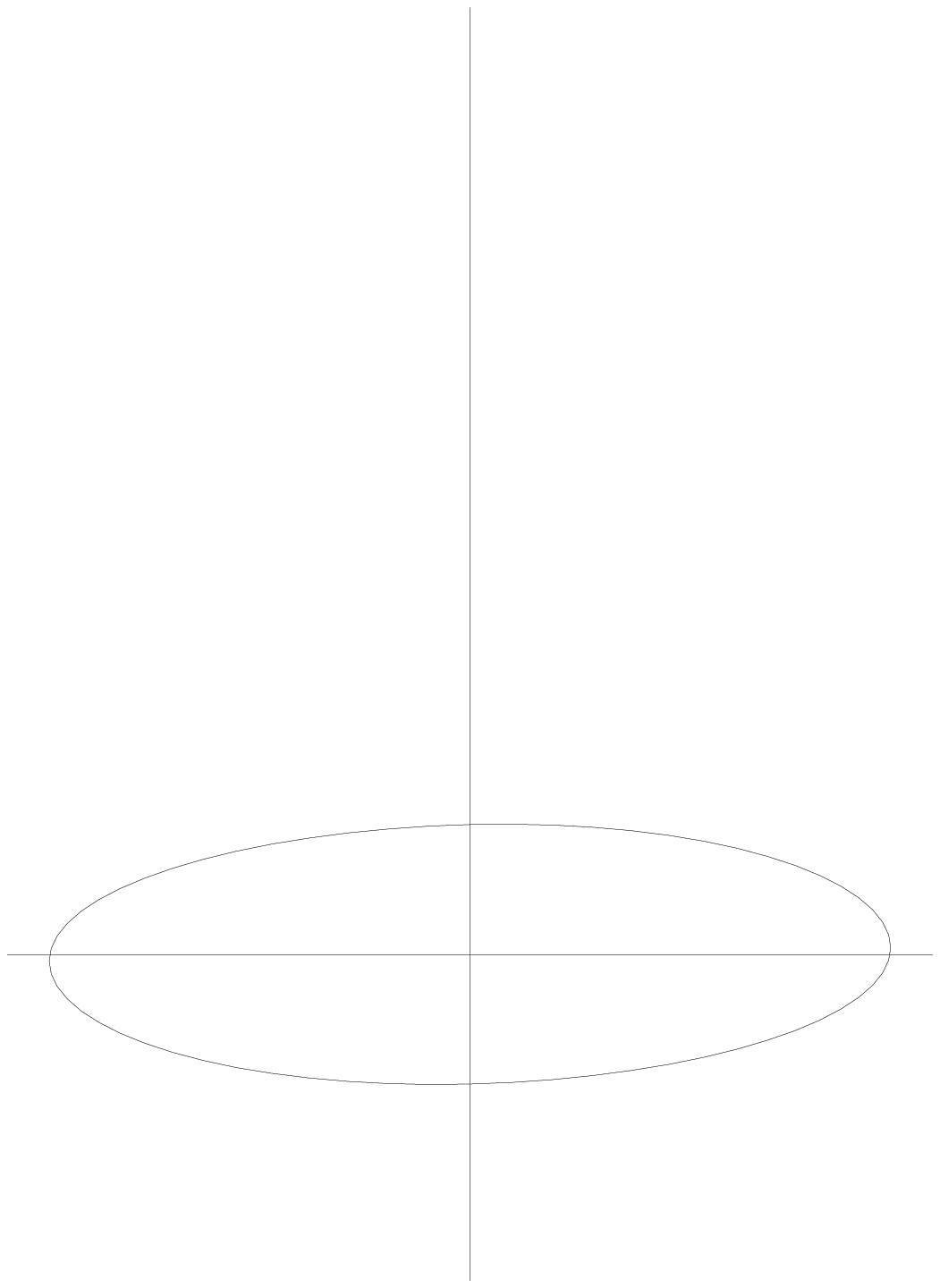}
\includegraphics{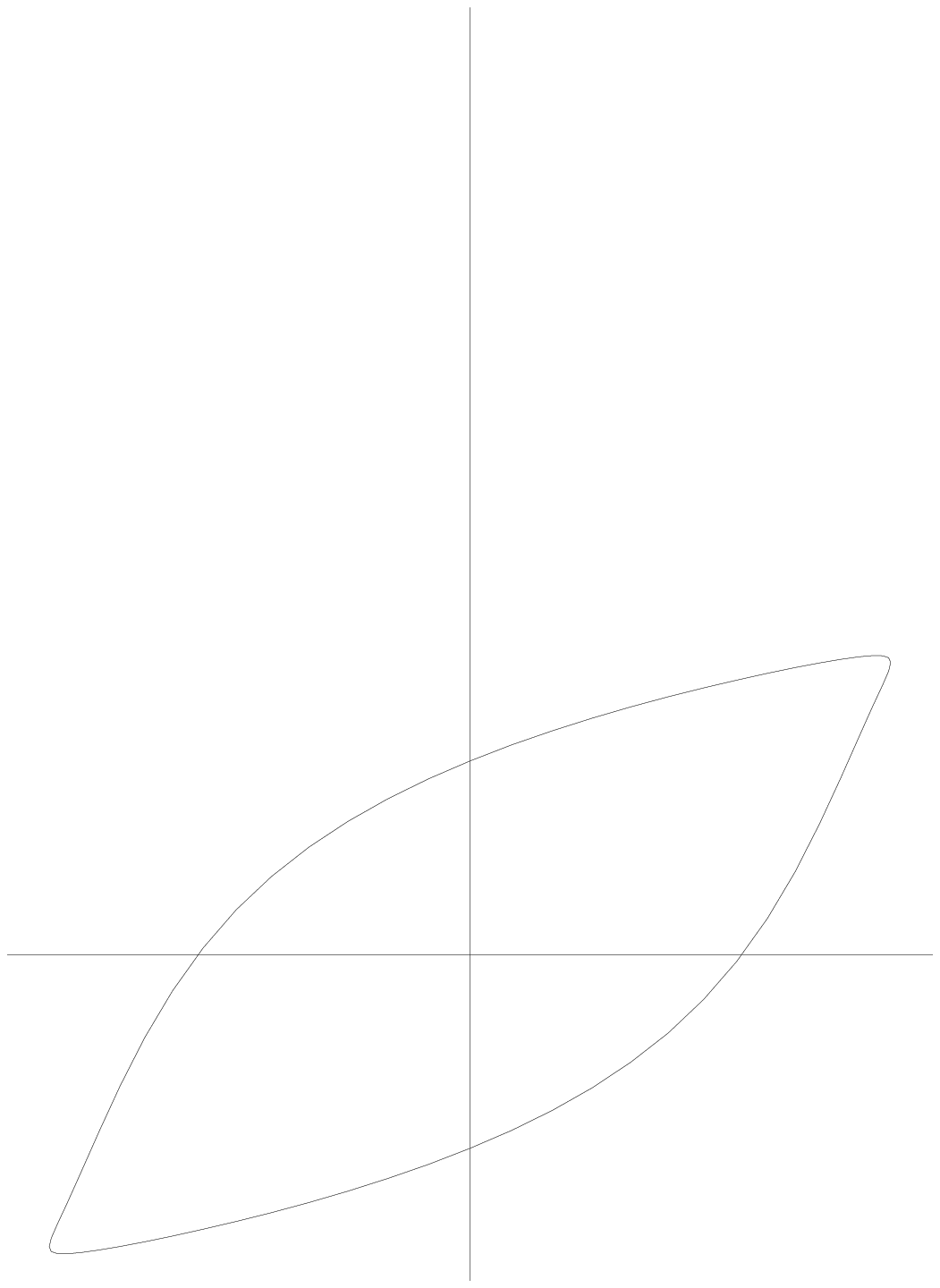}
\includegraphics{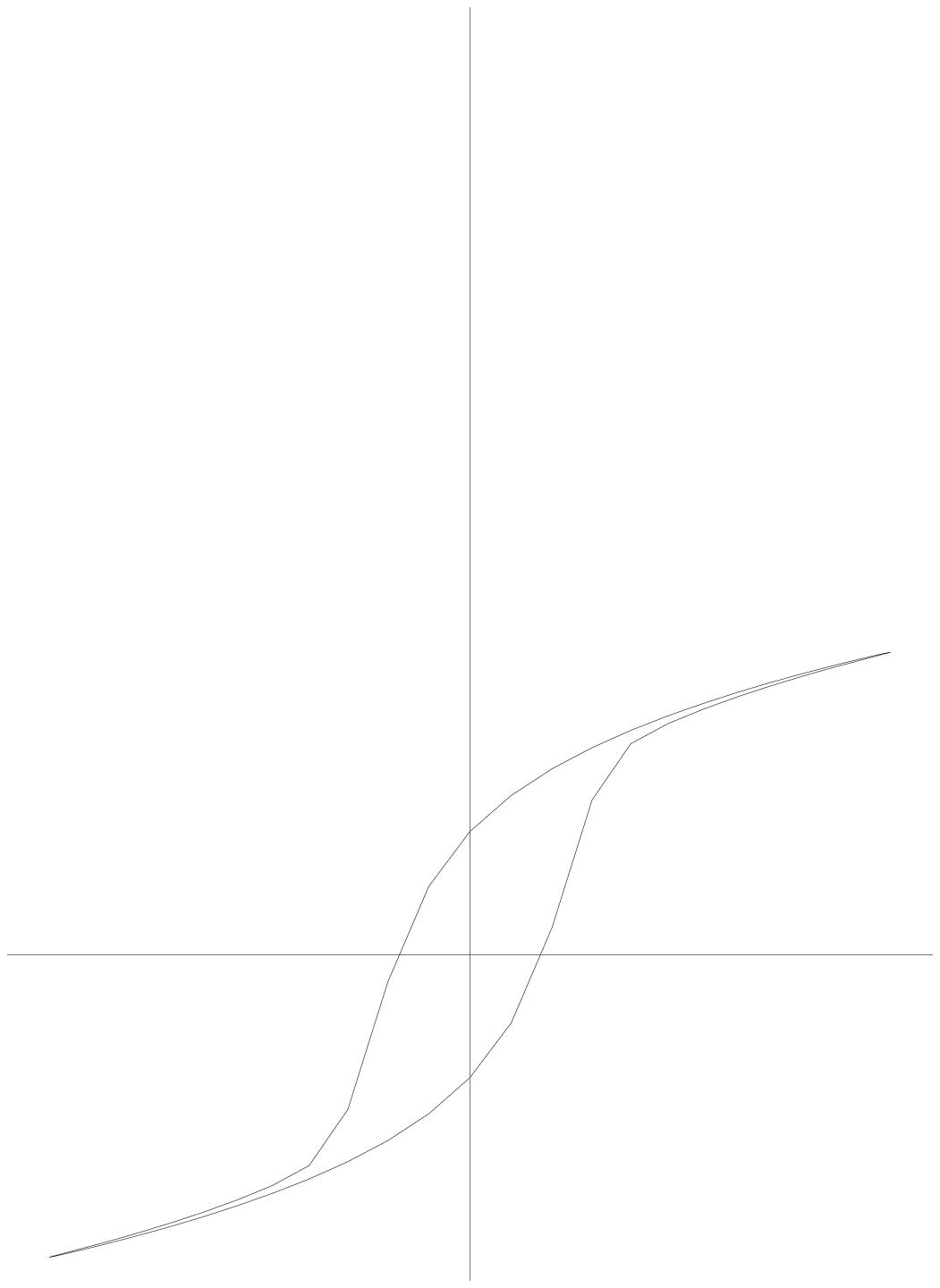}
The universal scaling function $F(\Omega)=h_0^{-\lambda} A(\omega h_0^\mu)$
describing hysteresis in the time-dependent Landau-Ginzburg theory. The
result for the critical $\phi^4$ theory is shown with a full line and that
for the tricritical $\phi^6$ theory with a dashed line. The universal loop
shapes $L(h/h_0;\Omega)$ in the $\phi^4$ model are shown below the main figure.
\endinsert
\midinsert\vskip13truecm
\centerline{FIGURE \figtag{ising}}
\includegraphics{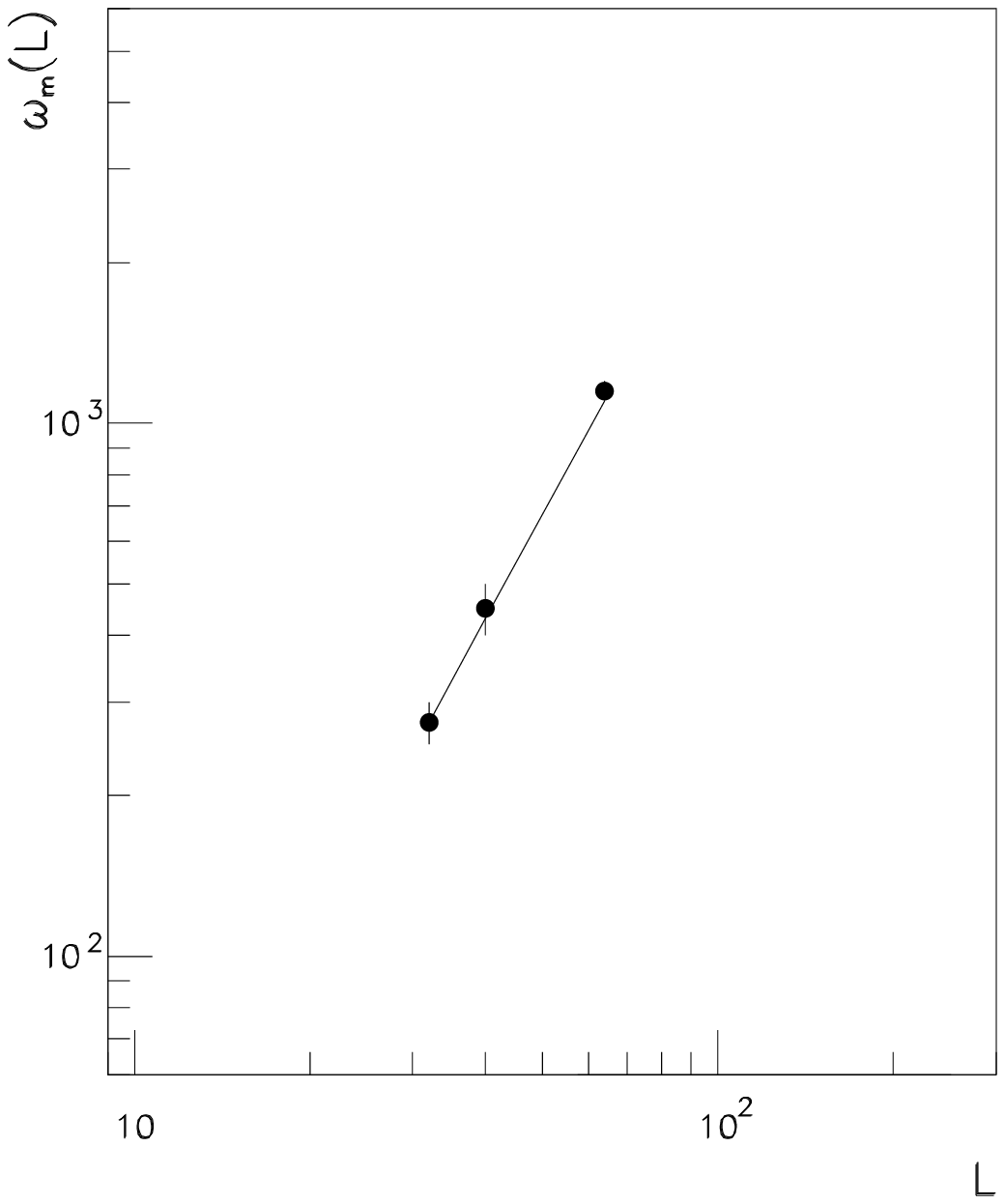}
Finite-size scaling tests for hysteresis in the two-dimensional Ising model.
The data have been obtained at couplings corresponding to the maximum of the
specific heat. The line shows the FSS prediction using the dynamical exponent
$z=2$.
\endinsert
\vfil\end